\documentclass[aps, superscriptaddress, floatfix, notitlepage, twocolumn]{revtex4-2}
\usepackage{epsfig}
\usepackage{graphicx}
\usepackage{amsmath}
\usepackage{bbold}
\usepackage{tabularx, booktabs}
\usepackage[normalem]{ulem}
\usepackage{xcolor}
\usepackage{multirow}
\usepackage{dutchcal} 
\usepackage[T1]{fontenc}
\usepackage{longtable}
\usepackage{appendix}
\usepackage[colorlinks=true,allcolors=black]{hyperref}
\usepackage{chemformula}
\let\ce\ch
\usepackage{orcidlink}

\usepackage{xr}
\usepackage{lineno}

\usepackage{float}
\usepackage[caption = false]{subfig}
\newcolumntype{Y}{>{\centering\arraybackslash}X}
\newcolumntype{Z}{>{\hsize=1.1\hsize\centering\arraybackslash}X}

\newcommand{\tc}{$T_{\rm c}$}

\begin{document}
\newcommand{\bochum}{Research Center Future Energy Materials and Systems of the University Alliance Ruhr and Interdisciplinary Centre for Advanced Materials Simulation, Ruhr University Bochum, Universitätsstraße 150, D-44801 Bochum, Germany}
\newcommand{\coimbra}{CFisUC, Department of Physics, University of Coimbra, Rua Larga, 3004-516 Coimbra, Portugal}
\newcommand{\mpi}{Max-Planck-Institut f\"ur Mikrostrukturphysik, Weinberg 2, D-06120 Halle, Germany}
\newcommand{\mlu}{Institut f\"ur Physik, Martin-Luther-Universit\"at Halle-Wittenberg, D-06099 Halle, Germany}
\newcommand{\upv}{Fisika Aplikatua Saila, Gipuzkoako Ingeniaritza Eskola, University of the Basque Country (UPV/EHU), Europa Plaza 1, 20018 Donostia/San Sebastián, Spain}
\newcommand{\cfm}{Centro de Física de Materiales (CFM-MPC), CSIC-UPV/EHU, Manuel de Lardizabal Pasealekua 5, 20018 Donostia/San Sebastián, Spain}
\newcommand{\dipc}{Donostia International Physics Center (DIPC), Manuel de Lardizabal Pasealekua 4, 20018 Donostia/San Sebastián, Spain}
\newcommand{\gdp}{Google DeepMind, Mountain View, CA, USA.}

\title{
Superconductivity in barium hydrides via incorporation of light elements
}


\author{Yue-Wen Fang\orcidlink{0000-0003-3674-7352}}
\email{yuewen.fang@ehu.eus}
\affiliation{\cfm}
\author{Ion Errea\orcidlink{0000-0002-5719-6580}}
\email{ion.errea@ehu.eus}
\affiliation{\upv}
\affiliation{\cfm}
\affiliation{\dipc}

\begin{abstract}
Barium hydrides are of interest for their potential in both ionic conductivity and superconductivity. Recently, a superconducting hydride \ce{BaH12} containing H$_2$ and H${_3}^{-1}$ molecular units was experimentally reported with a critical temperature $T_\text{c}$ of 20 K at 140 GPa [Nat Commun 12, 273 (2021)]. Herein, we combine ab initio methods with a rapid calculator of $T_\text{c}$ based on the networking value model to predict that the introduction of light elements, such as Be, can effectively expand the structure diversity and structure space of barium hydrides. 
Although molecular hydrogen units are still widely present in thermodynamically stable and metastable crystal structures, we find that a metastable phase of BeBaH$_8$ shows a high $T_\text{c}$ of 49 K at 100 GPa, which is only 
38 meV/atom above the thermodynamic stability energy.
This \ce{BeBaH8} remains dynamically stable at 15 GPa. Furthermore, our study shows that increasing pressure can further elevate $T_\text{c}$ beyond 100 K by enhancing the electron-phonon coupling constant. Our study proposes a feasible method for broadening the structural landscape in the exploration of superconducting phases of barium hydrides.

\end{abstract}
\maketitle

\clearpage


\section{Introduction}

In recent years, the exploration of hydrides has gained significant attention due to the pursuit of high critical temperature ($T_\text{c}$) superconductivity~\cite{kungao2025maximumtc,Lilia_2022-roadmap-JPCM,WendiZHAO-NSR2023,YingSUN-NSR2023,AFM2024-Tiago-hydrides-space}.  Both theoretical~\cite{fang-LuNH-CSP2024,sanna2023-Mg2IrH6,Sun-Li2MgH16-PRL2019,Errea2020-LaH10Nature,LaMgH-MaterTodayPhy2024,PRM2023-Superhydra,Duan2015-H4S2H2-SciRep} and experimental studies~\cite{Drozdov2015-H3S-Nature, PRL2023-LaBeH8-YanmingMA,Troyan2021-YH6-AdvMater} have suggested that high-$T_\text{c}$ and even near room temperature superconductivity is possible in hydrides, despite the fact that the materials must be subject to an extremely high pressure, e.g. the clathrate superhydride LaH$_{10}$ exhibits a $T_\text{c}$ of around 250 K at 175 GPa~\cite{LaH10-PRL2019-Somayazulu,Drozdov2019-La-H-Nature,Liu2017-LaH-PNAS}.
Because the application of extremely high pressure favors isotropic properties, the experimentally observed high-pressure superconducting hydrides usually feature very high-symmetry structures~\cite{Belli2021-network-value-NC,Lilia_2022-roadmap-JPCM}.

The neighbor of La in the periodic table, Ba, has been thought to be one of the promising host elements for the synthesis of superconducting hydrides~\cite{Chen2021-NatCommun-BaH12}. At ambient pressure, the $Pnma$ BaH$_2$ (\ce{BaH2}-I) is the thermodynamical stable phase of barium hydride. Heating at ambient pressure can drive a structural transition of BaH$_2$ from the low-symmetry $Pnma$ to the high-symmetry Ni$_2$In-type structure with the space group ${P6_3/mmc}$ (\ce{BaH2}-II). This structure transition makes BaH$_2$ be one of the prominent ionic conductors with potential applications in energy storage because the high-symmetry phase exhibits a hydrogen ion conductivity of 0.2 S$\cdot$cm$^{-1}$ at 630${^\circ}$C that is an order of magnitude larger than that of state-of-the-art oxide ion conductors~\cite{BaH2-2015Nature}.
However, both \ce{BaH2}-I and \ce{BaH2}-II are not good candidates for superconductivity because they are semiconductors with a wide band gap of around 1.9$\sim$2.9 eV~\cite{LU-BaH22007}.
The high-pressure experiments~\cite{Kinoshita2007-BaH12,Shuttleworth2023-BaH2}
suggest that \ce{BaH2} undergoes successive phase transformations: \ce{BaH2}-I first converts to \ce{BaH2}-II at around 2.5 GPa, then transforms to AlB$_2$-type simple hexagonal \ce{BaH2}-III accompanied by a closure of the band gap at around 57 GPa. Although \ce{BaH2}-III is metallic, superconductivity has not been observed. 
The experiments at high pressures also show that the BaH$_2$ can transform into BaH$_4$ by incorporating H$_2$ molecules at 38$\sim$40 GPa~\cite{Shuttleworth2023-BaH2,Jpclett-2022-Ba-H}. 
In the presence of molecular hydrogen in the synthesis setting, BaH$_4$ can be further compressed into Ba$_8$H$_{46}$ at around 45 GPa in which a Weaire-Phelan hydrogen network is formed. This hydrogen clathrate Ba$_8$H$_{46}$ shows semiconducting properties and can be stabilized down to 27 GPa~\cite{Jpclett2021-Ba8H46}.
It experiences a transition into a metallic state at around 85 GPa. However, both BaH$_4$ and Ba$_8$H$_{46}$ are not superconductors at these studied pressures~\cite{Shuttleworth2023-BaH2,Jpclett2021-Ba8H46,Jpclett-2022-Ba-H}.

Chen et al studied barium hydrides at higher pressures ranging from 75 to 173 GPa and reported several hydrogen-rich barium hydrides~\cite{Chen2021-NatCommun-BaH12}. In particular, their transport measurements revealed the onset of superconductivity at 20 K and 140 GPa in \ce{BaH12} with a pseudocubic face-centered cubic Ba sublattice. This superconducting hydride is suggested to be a unique molecular hydride because it contains H$_2$ and H$_3^{-1}$ molecular units, forming separate flat \ce{H12} chains, which are rarely observed in the high-symmetry clathrate superhydrides at high pressures.
In hydrides with H$_2$ and H${_3}^{-1}$ molecular units, the H-H bond disrupts the delocalized metallic bonding network, and their covalent nature may reduce the interaction between the vibrational modes of hydrogen atoms and the electronic states of the host lattice, which can lead to weaker electron-phonon coupling. Several recent studies have shown that the presence of hydrogen molecules generally form more insulating states than metallic states~\cite{fang-LuNH-CSP2024,2023arxiv2304.04447-Lilia-Chris}. Therefore, the molecular hydrogen units are generally not beneficial for superconductivity and high $T_\text{c}$.

Based on the networking value model and the statistics of hundreds of superconducting hydrides, an earlier study by one of the authors suggests that stretching the H-H bond or elimination of the hydrogen molecular units can favor  superconductivity~\cite{Belli2021-network-value-NC}.
Guided by this quantitative rule, in our study, we design new superconducting ternary barium hydrides at high pressures.
Aiming at eliminating or reducing the number of H$_2$ and H${_3}^{-1}$ molecular units in barium hydride, we introduce four elements with light atomic mass (Li, Be, B and C) to barium hydride and perform high-throughput crystal structure screening to locate the low-enthalpy structures up to 200 GPa. By studying the structural and electronic properties of a series of low-lying ternary barium hydrides, we find that H$_2$ molecular units remain widely observed in these ternary barium hydrides, while H${_3}^{-1}$ is almost absent. Due to the universal existence of molecular units in the studied ternary hydrides, more insulating states than metallic states are found in the low-enthalpy structure space.
Although metallic states were much fewer than the semiconducting or insulating phases, there are several metastable metallic molecular hydrides showing interesting properties.
In particular, \ce{BeBaH4} with tetragonal structure is nearly thermodynamically stable at 100 GPa and can be still dynamically stable at 50 GPa.
In addition, by combining the quick estimator of superconducting $T_\text{c}$ based on the networking value model~\cite{Belli2021-network-value-NC,Network-Trinidad2024} with the ab initio methods, \ce{BeBaH8} with orthorhombic structure has been predicted to be superconducting with an estimated $T_\text{c}$ of 49 K at 100 GPa.
The value of $T_\text{c}$ can be further boosted by increasing the pressure, reaching 107 K at 200 GPa.
This \ce{BeBaH8} remains dynamically stable down to 15 GPa although it is not superconducting due to the substantial reduction of the electron-phonon coupling constant with pressure lowering.

\section{Methods}

\subsection{Crystal structure prediction}

Crystal structure prediction methods, i.e., the particle swarm algorithm implemented in CALYPSO~\cite{Wang2010,Wang2012} and the evolutionary algorithm implemented in CrySPY~\cite{Yamashita2021CrySPY}, were used to predict the crystal structures of $A$-Ba-H ($A$ = Li, Be, B and C) at pressures up to 200 GPa.
The crystal structure prediction was performed with a fixed composition, and cell sizes were explored up to 32 atoms per cell. A total of around 100,000 structures were screened during the crystal structure prediction.
The enthalpy and forces of the predicted crystal structures were calculated by first-principles density functional theory (DFT) methods. The DFT calculations were carried out using the Vienna Ab initio Simulation Package (VASP)~\cite{Kresse-PRB-1996,Kresse1996} employing the projector-augmented wave (PAW) method. The exchange-correlation functional was treated in the generalized gradient approximation within the parameterization of Perdew, Burke, and Ernzerhof~\cite{PBE-Perdew1996}.
To generate $k$-points for different crystal structures during structure screening, we employed the $k$-point generation scheme in Pymatgen~\cite{Jain2011_pymatgen}, using a grid density of 100 $k$-points per \AA$^{-3}$ of reciprocal cell volume. An energy cutoff of 450 eV was used in the crystal structure prediction. To construct the convex hull phase diagram,  we considered all materials available in our crystal structure prediction, as well as those we could find from the Materials Project~\cite{Jain2013-MaterProject}. 
The structures were fully optimized until the energy convergence criterion of 10${^{-8}}$ eV and force criterion of 10${^{-3}}$ eV${\cdot}$\AA$^{-1}$ were satisfied. 

\subsection{Electronic structure calculations}

In high-throughput DFT calculations of the density of states (DOS), we used the Gaussian smearing method in VASP. To ensure accurate results, we used a narrow smearing width of 0.05 eV accompanied by a dense $k$ point grid with 200 points per \AA$^{-3}$ of reciprocal cell volume.
To extract the total DOS and the element/orbital projected DOS, we used the Sumo~\cite{Ganose2018-sumocode} code. In both band and phonon dispersion calculations, the special $k$-path were generated using Sumo~\cite{Ganose2018-sumocode}. To estimate the superconducting critical temperature \tc~based on the networking value model, we used the TcESTIME code~\cite{Belli2021-network-value-NC,Network-Trinidad2024}. The value of $T_\text{c}$ with an error of 60 K is correlated to $\Phi_{\text{DOS}}$:
\begin{equation}
    T_c = (750 \Phi_{\text{DOS}} - 85) \, \text{K},
\end{equation}
\begin{equation}
    \Phi_{\text{DOS}} = \phi H_f {H_{\text{DOS}}^{-3}} 
\end{equation}
where $\phi$ is the networking value based on the electron localization function, $H_f$ is the hydrogen fraction, and $H_{\text{DOS}}$ is the hydrogen fraction of the total DOS at the Fermi energy.

\subsection{Phonon and electron-phonon coupling calculations}

The phonon, and electron-phonon coupling properties were investigated using density-functional perturbation theory (DFPT) method implemented in Quantum Espresso~\cite{QE-2009,QE-2017}. The dynamical stability was also cross-checked by the supercell and finite displacement methods implemented in Phonopy~\cite{Togo-phonopy2015}. 
The Quantum Espresso calculations were performed using norm-conserving pseudopotentials from the strict set of PseudoDojo~\cite{vanSetten2018pseudodojo}.
In Quantum Espresso calculations, geometry optimizations were carried out using uniform $\Gamma$-centered $k$-point grids with a density of 3000 $k$-points per \AA$^{-3}$.
In the electron-phonon coupling calculations, the $k$-grid from the structure relaxation was doubled, and even quadrupled for cross-check in some cases, in each direction. In the particular case of \ce{BeBaH8}, the results converged well with a $k$-grid of ${42 \times 42 \times 42}$. The critical temperature for superconductivity was calculated using the Allen-Dynes formula:
\begin{equation}
T_c = \frac{f_1 f_2 \omega_{\text{log}}}{1.20} \exp \left( \frac{-1.04 (1 + \lambda)}{\lambda - \mu^* (1 + 0.62 \lambda)} \right),
\end{equation}
\begin{equation}
f_1 = \left( 1 + \left( \frac{\lambda}{2.46 (1 + 3.8 \mu^*)} \right)^{3/2} \right)^{1/3},
\end{equation}
\begin{equation}
f_2 = \left( 1 + 
\frac{\lambda^2 \left( \frac{\bar{\omega}_2}{\omega_{\text{log}}} - 1 \right)}
{\lambda^2 + [1.82 (1 + 6.3 \mu^*) (\frac{\bar{\omega}_2}{\omega_{\text{log}}})]^2}
\right),
\end{equation}
where \( f_1 \) and \( f_2 \) are correction factor depending on the electron-coupling constant \( \lambda \),  Coulomb pseudopotential parameter \( \mu^* \), average logarithm frequency \( \omega_{\text{log}} \), and \( \bar{\omega}_2 \). The frequencies \( \bar{\omega}_n \) are the \( n^{\text{th}} \) root of the \( n^{\text{th}} \) moment of the normalized distribution ${g(\omega) = \frac{2}{\lambda \omega} \alpha^2 F(\omega)}$~\cite{ML-Eliashberg2022npj}. In our study, the Coulomb pseudopotential parameter (${\mu^*}$) of 0.1 was used.

\section{Results}

\subsection{$A_2$BaH$_{12}$ at 200 GPa where $A$ = Li, Be, B, and C}

\begin{figure}[hpt]
\centering
\includegraphics[angle=0,width=0.5\textwidth]{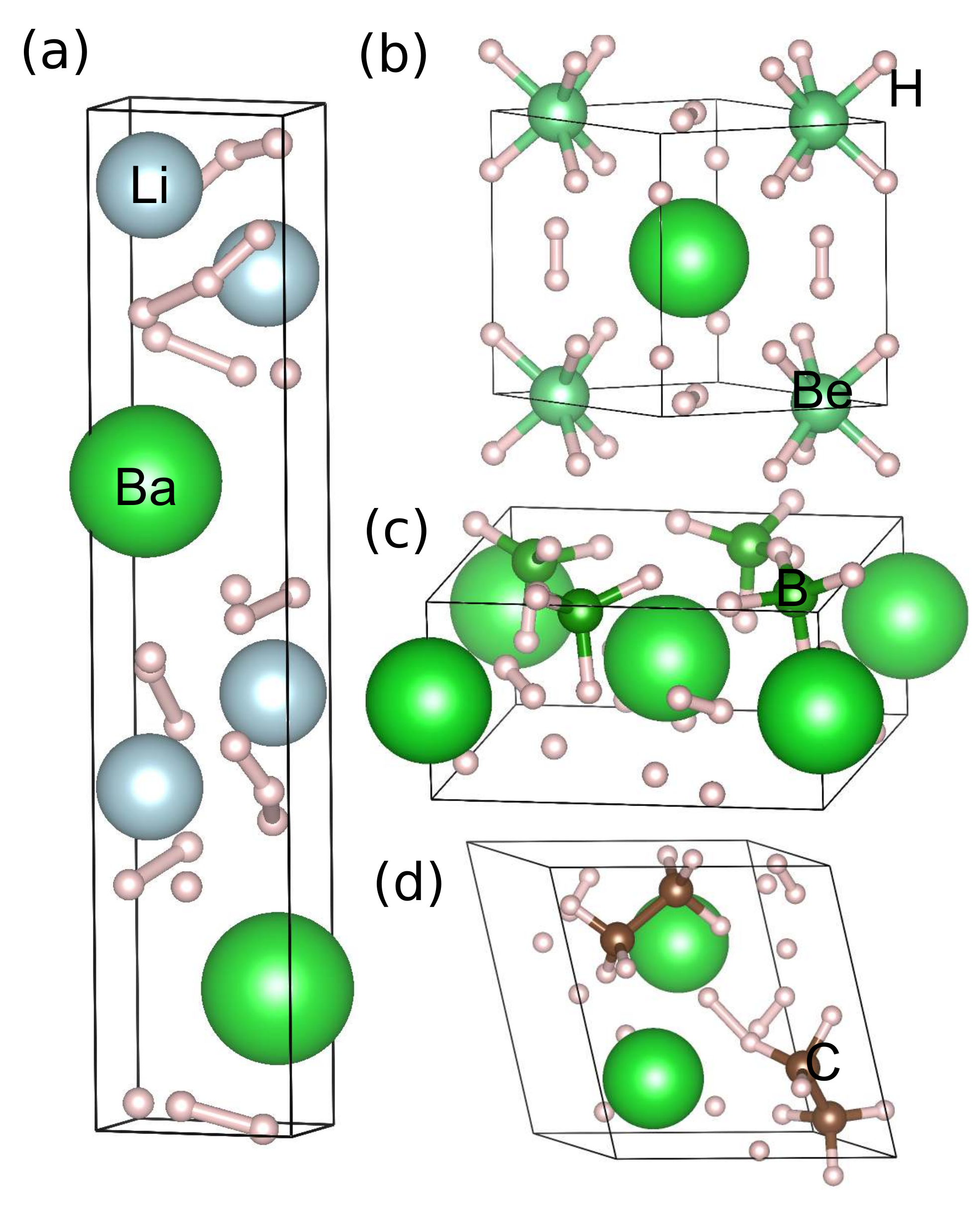}
\caption{\label{fig:A2BaH12-200GPa-GS-strcture}  
The lowest-lying states at 200 GPa of (a) Li$_2$BaH$_{12}$ with space group $Cc$,  (b) Be$_2$BaH$_{12}$ with space group $Cmmm$,  (c) B$_2$BaH$_{12}$ with space group $P4bm$, and (d) C$_2$BaH$_{12}$ with $P1$. The H-H bond is plotted if the distances are less than 1.0 \AA,  $A$-$A$ and $A$-H bonds are plotted if the distances are less than 1.4 \AA.
}
\end{figure}

Chen \textit{et al.}~studied Ba-H compounds up to 200 GPa and found that the molecular hydride BaH$_{12}$, with H$_2$ and H$_3^{-1}$ molecular units, exhibited a $T_{\rm c}$~of 20 K at 140 GPa~\cite{Chen2021-NatCommun-BaH12}. 
To investigate if the introduced light-mass elements ($A$ = Li, Be, B, and C) can break the H$_2$ and H$_3^{-1}$ molecular units and form more metallic phases at high pressures, as well as to check the chemical compatibility of the introduced elements, we first performed crystal structure predictions for the two formula units $A_2$BaH$_{12}$ (i.e. 30-atom A$_4$Ba$_2$H$_{24}$) at 200 GPa as a test case.
In each crystal structure prediction, we carried out around 20-50 generations for each fixed stoichiometry, with each generation including 100 crystal structures.  For each stoichiometry of $A_2$BaH$_{12}$, the structural and electronic properties of the twenty low-lying states are studied. The space group and enthalpy of the low-lying structures for \ce{LiBaH12}, \ce{BeBaH12}, \ce{BBaH12} and \ce{CBaH12} are shown in Supplementary Tables I-IV. The crystal structures of the lowest-enthalpy states at 200 GPa, as predicted by our crystal structure calculations, are shown in Figure~\ref{fig:A2BaH12-200GPa-GS-strcture}.

Because the 2 f.u. $A_2$BaH$_{12}$ is a large cell and most predicted structures show low symmetry (most of them being $P1$ without any symmetry operation except the identity operation), it is computationally demanding to carry out a high-throughput calculation of $T_\text{c}$ using DFPT methods as implemented in Quantum Espresso. Alternatively, we note that the networking value model~\cite{Belli2021-network-value-NC,Network-Trinidad2024} has been widely recognized and used in many recent studies to predict superconducting critical temperatures in hydrogen-based superconductors~\cite{Adam-PRMat2024-LuNh,MaterTodayPhy2024-ML-superconductors,Adam-design-quaternary2024-PNAS}. Because the networking value model is based on electron localization functions that can be obtained in standard ab initio calculations, it makes the estimation of \tc~much computationally cheaper compared to the direct computation from solving Eliashberg spectral functions in DFPT calculations. 
Previously, the networking value model has been applied in the high-throughput ab initio screening of superconducting hydrides, accelerating the theoretical identification of promising high-\tc~candidates. With the aid of the networking value model, our previous DFPT calculations predict Lu$_4$H$_{11}$N to have a \tc~of 100 K at 20 GPa, and binary LuH$_6$ and LuH$_{10}$ to be near-room-temperature superconductors at pressures above 100 GPa~\cite{fang-LuNH-CSP2024}.

Using the networking value model method implemented in TcESTIME~\cite{Belli2021-network-value-NC,Network-Trinidad2024}, we first estimate the superconducting critical temperature of BaH$_{12}$ which has an experimental \tc~of 20 K at 140 GPa according to the report by Chen et al~\cite{Chen2021-NatCommun-BaH12}. 
Our calculation was based on the relaxed crystal structure proposed by Chen et al, i.e. ${Cmc2_1}$ BaH$_{12}$. Our networking value model yields a \tc~of 70 K in BaH$_{12}$ at 140 GPa.
In Chen et al's study, they reported a DFPT-calculated \tc~of 53 K. However, this ${Cmc2_1}$ BaH$_{12}$ in their ab initio calculations shows a number of imaginary modes, which indicates the instability, therefore, an accurate theoretical description of crystal structure and superconductivity of BaH$_{12}$ remains an open question. Considering the networking value model has an internal error of ${\pm 60}$ K~\cite{Belli2021-network-value-NC,Network-Trinidad2024}, the networking value model can work as a useful indicator to screen promising superconducting hydrides. Next, this method is used to compute \tc~for the low-lying metallic structures of the 30-atom $A_2$BaH$_{12}$ at 200 GPa. The values obtained are shown in the Supplementary Tables~I$\sim$IV. The highest estimated \tc~for \ce{Li2BaH12} is 56 K, while no superconducting phase is observed in \ce{C2BaH12}. Both \ce{BeBaH12} and \ce{BBaH12} show estimated critical temperatures above 70 K which approach the boiling point of liquid nitrogen. 

In examining B$_2$BaH$_{12}$ and C$_2$BaH$_{12}$, we note from Supplementary Tables I$\sim$IV that more than half of the structures among the twenty low-enthalpy states exhibit insulating character.
Relatively, more low-lying structures of Li$_2$BaH$_{12}$ and Be$_2$BaH$_{12}$ show metallic conductivity.
The analysis of these low-lying structures of $A$BaH$_{12}$ at 200 GPa demonstrates that the introduction of Li and Be into the barium hydrides has more possibilities to form metallic systems than B and C. 
In particular, the lowest-lying structures of $P4bm$ B$_2$BaH$_{12}$ and $P1$ C$_2$BaH$_{12}$,  illustrated in in Figure~\ref{fig:A2BaH12-200GPa-GS-strcture} are both insulating, whereas the lowest-lying states of $Cc$ Li$_2$BaH$_{12}$ and $Cmmm$ Be$_2$BaH$_{12}$ are both metallic. 
Although the lowest-lying structures of \ce{Li2BaH12} and \ce{Be2BaH12} are metals, the minimal interatomic distances in Table~\ref{tab:A2BaH12-atomic-dis} show that their minimal H-H bond lengths are 0.830 and 0.756 \AA, respectively, which are just slightly larger than the minimal H-H bond lengths in \ce{B2BaH12} and \ce{C2BaH12}. Therefore, the H$_2$ molecule units are present in all the ground-state structures of $A$BaH$_{12}$ (see Figure~\ref{fig:A2BaH12-200GPa-GS-strcture}). 
However, unlike the other three cases but like the ${Cmc2_1}$ \ce{BaH12} reported by Chen et al~\cite{Chen2021-NatCommun-BaH12}, the H-H-H (H${_3}^{-1}$) chain is only observed in \ce{Li2BaH12}, as illustrated in Figure~\ref{fig:A2BaH12-200GPa-GS-strcture}(a). 

\begin{table}[h]
\centering
\caption{The minimal interatomic distances in the lowest-enthalpy states of $A$BaH$_{12}$ at 200 GPa, where $A$ = Li, Be, B, and C.}
\label{tab:A2BaH12-atomic-dis}
\begin{tabular}{ccccc}
\hline
Element $A$ & H-H (\AA) & $A$-H (\AA) & $A$-$A$ (\AA) & Ba-H (\AA)  \\
\hline
Li & 0.830 & 1.402 & 1.855 & 1.971 \\
Be & 0.756 & 1.350 & 4.011 & 2.066 \\
B & 0.741 & 1.100 & 3.108 & 2.069 \\
C & 0.747 & 1.046 & 1.377 & 1.975 \\
\hline
\end{tabular}
\end{table}

\subsection{BeBaH$_{x}$ at pressures}

Because the extremely high pressure up to 200 GPa reported in Chen et al's study~\cite{Chen2021-NatCommun-BaH12} is not feasible in most laboratory settings, we were motivated to search for superconducting phases that could be stabilized at or below one megabar. To this end, we shifted our focus onto the exploration of the potential superconducting structures among the structural phase space of $A$-Ba-H at 100~GPa.
Given that the analysis at 200 GPa indicates that the incorporation of Be and Li into Ba-H promotes the formation of more metallic states compared to B and C, and that the upper limit of \tc~is expected to be higher for Be doped compounds than for Li (see Supplementary Tables I and II), we further limit our search scope in BeBaH$_{x}$ ($x = 2n, \quad n \in \{1, 2, 3, \dots, 12\}$). 
Figure~\ref{fig:BeBaHx-100GPa-hull} shows the phase diagram at 100 GPa based on the crystal structures from our crystal structure prediction, Materials Project~\cite{Jain2013-MaterProject} an other dataset in literature~\cite{Chen2021-NatCommun-BaH12,Jpclett2021-Ba8H46,Shuttleworth2023-BaH2,Yu-AIP-Adv-2014-BeH2}.
We find that ${P\bar{6}m2}$ BeBaH$_6$ is on the convex hull, which implies that it is thermodynamically stable. However, it is not a superconductor. The crystal structure of BeBaH$_6$ is shown in Fig.~\ref{fig:BeBaHx-100GPa-hull} and is more specifically shown in Supplementary Figure 1.

\begin{figure}[hpt]
\centering
\includegraphics[angle=0,width=0.5\textwidth]{./img/100GPa-phase-diagram-v2.2.pdf}
\caption{\label{fig:BeBaHx-100GPa-hull}  
The phase diagram at 100 GPa along with the crystal structures of  thermodynamically stable BeBaH$_6$, and two metastable phases including \ce{BeBaH4} and \ce{BeBaH8}.
The structural parameters and atomic coordinates of \ce{BeBaH4}, \ce{BeBaH6}, and \ce{BeBaH8} are listed in Supplementary Table V of Supplementary Material.
}
\end{figure}

Next we study the metallic stoichiometries lying within 100 meV/atom from the convex hull, which is a good reference for potential metastability. In particular, by estimating $T_\text{c}$ using the networking value model, we find that two hydrides, BeBaH$_4$ and BeBaH$_8$, are potential superconductors with $T_\text{c}$  values being 25 K ($\pm$ 60 K) and 65 K ($\pm$ 60 K), respectively.  These two hydrides BeBaH$_4$ and BeBaH$_8$ are 11 meV/atom and 38 meV/atom above the convex hull, respectively at 100 GPa. Since these two compounds are not far from the convex hull, implying the feasibility of experimental synthesis, the electron-phonon coupling properties of these two structures were further investigated using DFPT method.

The hydride \ce{BeBaH4} has a tetragonal crystal structure and a space group of ${P4/mmm}$ (No. 123) at 100 GPa, in which $a$ = 2.91 \AA~and $c$ = 3.87 \AA. The crystal structure is shown in Fig.~\ref{fig:BeBaHx-100GPa-hull}. 
Its electron localization function (Supplementary Figure 2 in Supplementary Material) suggests that \ce{BeBaH4} is an ionic system.
At 100 GPa, the phonon spectrum in Fig.~\ref{fig:BeBaH4-100GPa-ph} shows that this ${P4/mmm}$ \ce{BeBaH4} is dynamically stable.  However, because of the small electron-phonon coupling constant $\lambda$ of 0.15, the Allen-Dynes formula predicts that it is not a superconductor at 100 GPa. We find that a higher pressure of 200 GPa does not help to develop superconductivity because the value of $\lambda$ is only slightly increased to 0.16. Despite the absence of superconductivity, this tetragonal \ce{BeBaH4} can be stabilized at a lower pressure of 50 GPa.  
By comparing the phonon spectra at 50 and 100 GPa, we can find that the decreased pressure only has a small effect on the low-energy phonon modes consisting of heavy atoms Be and Ba, but in the meantime it softens the high-energy hydrogen modes significantly. 

\begin{figure}[htbp]
\centering
\includegraphics[angle=0,width=0.49\textwidth]{./img/BeBaH4-v2}
\caption{\label{fig:BeBaH4-100GPa-ph}  
The electronic and phonon properties of \ce{BeBaH4} at 100 and 50 GPa. (a) Electronic band structure, electronic density of states (DOS), phonon band structure and phonon density of states (PDOS) at 100 GPa. (b) Electronic band structure, DOS, phonon band structure and PDOS at 50 GPa.
}
\end{figure}

The crystal structure of \ce{BeBaH8} has an orthorhombic Bravais lattice and a space group of ${Cmc2_1}$ (No. 36). There are two formula units in the primitive cell, resulting in 20 atoms in the primitive cell. The primitive cell of \ce{BeBaH8} is shown schematically in Fig.~\ref{fig:BeBaHx-100GPa-hull}.
At 100 GPa, the electronic band structure and the density of states in Fig.~\ref{fig:BeBaH8-100GPa-ph} suggest that \ce{BeBaH8} is conductive. There are two bands crossing the Fermi level, forming one electron pocket at Y and one hole pocket at $\Gamma$. More importantly, the electronic states at the Fermi level are dominated by hydrogen atoms with minor contributions from the heavy host metals Ba and Be, signaling that \ce{BeBaH8} may be a hydride superconductor.
The phonon dispersions and phonon density of states (PDOS) do not show any imaginary mode, indicating that \ce{BeBaH8} is dynamically stable at 100 GPa. 
Furthermore, we find that \ce{BeBaH8} remains dynamically stable in the harmonic approximation at pressures down to only 15 GPa, as evidenced by the phonon bands and PDOS shown in Fig.~\ref{fig:BeBaH8-100GPa-ph}(b).
However, the electronic band structure and DOS in Fig.~\ref{fig:BeBaH8-100GPa-ph}(b) show that  \ce{BeBaH8} enters into the semiconducting state, exhibiting an indirect band gap of 1.5 eV at this lower pressures.
The evolution of electron localization functions of BeBaH$_8$ from 100 GPa down to 15 GPa is also shown in Supplementary Figure 3 in Supplementary Material.
Despite the gap opening at reduced pressures, our calculations suggest that the BeBaH$_8$ can be stabilized at a moderate pressure of only 15 GPa. It is worth noting that incorporating ionic quantum and anharmonic effects using the stochastic self-consistent harmonic approximation (SSCHA) method may help stabilize it even at ambient pressure~\cite{Errea-PRB2014-SSCHA, Monacelli-JPCM2021-SSCHA, RbPH3-2024-arxiv,fang-LuNH-CSP2024,dangic2023-arxiv-color-change-parent-structure}.

\begin{figure}[htbp]
\centering
\includegraphics[angle=0,width=0.49\textwidth]{./img/BeBaH8-phonon-v2.pdf}
\caption{\label{fig:BeBaH8-100GPa-ph}  
The electronic and phonon properties of \ce{BeBaH8} at 100 and 15 GPa. (a) Electronic band structure, electronic density of states (DOS), phonon band structure and phonon density of states (PDOS) at 100 GPa. (b) Electronic band structure, DOS, phonon band structure and PDOS at 15 GPa.
}
\end{figure}

\begin{figure}[htp]
\centering
\includegraphics[angle=0,width=0.49\textwidth]{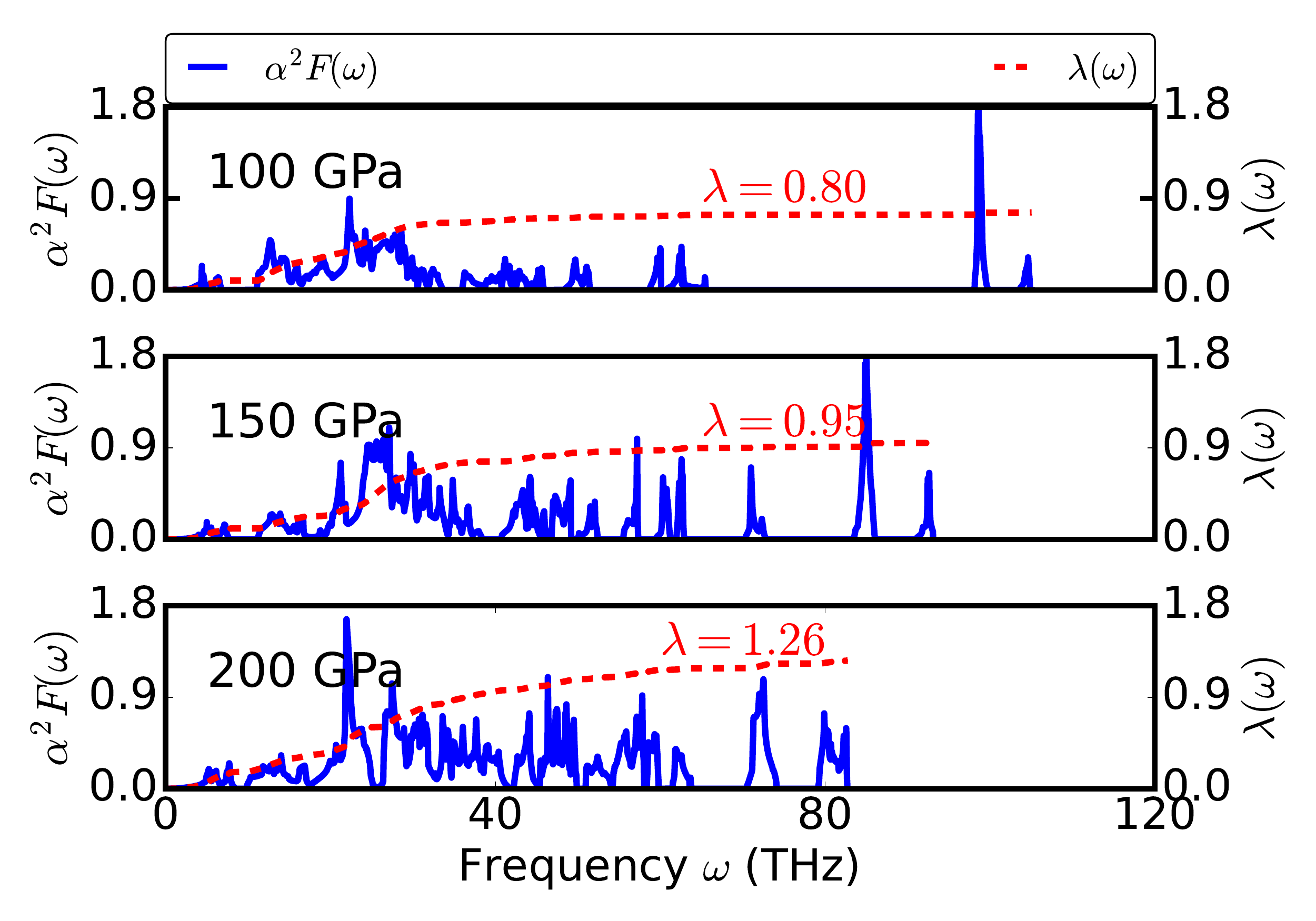}
\caption{\label{fig:BeBaH8-a2F}  
Eliashberg spectral function $\alpha^2F(\omega)$ (solid blue lines) and the cumulative frequency-dependent electron-phonon coupling function $\lambda$($\omega$) (dashed red lines) of
\ce{BeBaH8} at 100, 150 and 200 GPa.
}
\end{figure}

By solving the Eliashberg spectral function $\alpha^2F(\omega)$,
the electron-phonon coupling constant of BeBaH$_8$ is calculated to be around 0.8 at 100 GPa. The corresponding Eliashberg spectral function $\alpha^2F(\omega)$ and the cumulative frequency-dependent electron-phonon function $\lambda({\omega)}$ are shown in Fig.~\ref{fig:BeBaH8-a2F}. By comparing the $\alpha^2F(\omega)$ and the phonon density of states in Fig.~\ref{fig:BeBaH8-100GPa-ph}, we find that the electron-phonon coupling is dominated by the phonon modes of H and Be character ranging from 10 THz to 40 THz, followed by the minor contribution of phonon modes of Ba below 10 THz.

Using Allen-Dynes equation with Coulomb pseudopotential parameter (${\mu^*}$) of a typical value 0.1, $T_\text{c}$ of BeBaH$_8$ is calculated to be around 49 K. This obtained $T_\text{c}$ from Quantum Espresso calculations is comparable to that (65 $\pm$ 60 K) predicted by the networking value model, showing its validity to screen high-\tc\ compounds. 
Upon increasing pressure, we find that the electron-phonon coupling can be enhanced significantly.
The Eliashberg spectral functions $\alpha^2F(\omega)$ of 150 GPa and 200 GPa are compared to that at 100 GPa in Fig.~\ref{fig:BeBaH8-a2F}. Due to the overall softening of the high-frequency pure H modes, these modes join to contribute additional electron-phonon coupling at 150 and 200 GPa, compared to the case of 100 GPa in which the high-frequency pure H modes almost have negligible contributions to $\lambda$. As a result, the values of $T_\text{c}$ are increased to 73 K and 107 K at 150 GPa and 200 GPa, respectively.

\section{Discussion \& Conclusions}

By combining the quick estimator of superconducting $T_\text{c}$ with the high-throughput first-principles structure screening, we explore the possibility of implementation of superconductivity in the ternary barium hydrides $A$-Ba-H ($A$ = Li, Be, B and C). Although the introduction of these four light elements into barium hydrides does not eliminate the H$_2$ or H$_3^{-1}$ molecular units completely in the low-lying structure space, they have significantly changed the low-lying crystal structures compared to those in Ba-H.

Among the four $A$-Ba-H compositions, Be-Ba-H is found to be most promising to host superconductivity at pressures. Our calculations find that ${P\bar{6}m2}$ BeBaH$_6$ is the only phase lying on the convex hull among the Be-Ba-H compounds, suggesting that it is thermodynamically stable and can be synthesized in experiment. In addition, we also find two interesting metastable phases--BeBaH$_4$ and BeBaH$_8$, which 
lie only 11~and 38~meV/atom above the convex hull at 100~GPa, respectively. This indicates that BeBaH$_4$ and BeBaH$_8$ are thermodynamically close to stability and thus may be accessible experimentally, especially under high-pressure synthesis conditions. 
Because the distances to the convex hull in this work are calculated based on DFT Born-Oppenheimer energy calculations,  they could be significantly higher than the actual values, as was the case for LaH$_{10}$~\cite{Errea2016Quantum}. The more accurate distances to the convex hull can be addressed by using more advanced ab initio calculations such as stochastic self-consistent harmonic approximation (SSCHA) method that can incorporate ionic quantum and anharmonic effects~\cite{Errea-PRB2014-SSCHA, Monacelli-JPCM2021-SSCHA, RbPH3-2024-arxiv,fang-LuNH-CSP2024,dangic2023-arxiv-color-change-parent-structure}.

In spite of the internal error of $\pm$ 60 K in the networking value model implemented in TcESTIME~\cite{Belli2021-network-value-NC,Network-Trinidad2024}, this tool TcESTIME remains highly valuable for rapidly screening potential superconductors, offering a much more computationally efficient alternative to standard DFPT calculations.
In our study, we used DFPT calculations only for candidates with TcESTIME-predicted \tc~above 20 K. Given the model's error margin, this threshold ensures that we do not overlook promising superconducting phases, while still significantly reducing the computational cost. Therefore, our main results regarding the most promising compounds are robust with respect to the model's uncertainty.
In particular, our ab initio calculations predict a high $T_\text{c}$ of 49 K in BeBaH$_8$ at a pressure of 100 GPa. The superconductivity in \ce{BeBaH8} is effectively modulated by pressure, as the electron-phonon coupling constant is strongly affected by pressure. At higher pressures, such as 200 GPa, the critical temperature $T_\text{c}$ reaches 107 K, while at lower pressures around 15 GPa, superconductivity vanishes with the band gap opening. 

Our study demonstrates a method to expand the structural space of barium hydrides by incorporating light elements, thereby maintaining superconductivity at reduced pressures. This approach is expected to stimulate further research efforts in synthesis and experimental measurements of barium hydrides, and will be applicable to design other moderate or ambient pressure hydride superconductors.

\section{Acknowledgments}
We thank Wuhao Chen and Dmitrii V. Semenok for the helpful discussions.
This work is supported by the European Research Council (ERC) under the European Unions Horizon 2020 research and innovation program (Grant Agreement No. 802533), the Spanish Ministry of Science and Innovation (Grant No. PID2022142861NA-I00), the Department of Education, Universities and Research of the Eusko Jaurlaritza and the University of the Basque Country UPV/EHU (Grant No. IT1527-22), the IKUR Strategy under the collaboration agreement between Ikerbasque Foundation and DIPC on behalf of the Department of Education of the Basque Government, and Simons Foundation through the Collaboration on New Frontiers in Superconductivity. Y.-W.F. was also supported by Extraordinary Grant of CSIC (Grant No. 2025ICT122).
We acknowledge EuroHPC for granting us access to Lumi located in CSC’s data center in Kajaani, Finland, (Project ID EHPC-REG-2024R01-084) and to RES for giving us access to MareNostrum5, Spain, (Project ID FI-2024-2-0035). Technical and human support provided by DIPC Supercomputing Center
is gratefully acknowledged. The authors acknowledge enlightening discussions with the partners of the SuperC collaboration. 

\section{Data Availability Statement}
The data supporting this study's findings are available within the article.

\subsection{Authors’ contributions}
Y.-W.F. and I.E. designed this project and interpreted the data. 
Y.-W.F. performed the calculations and wrote the manuscript with input from I.E.











%
\end{document}